\documentclass{tinytocs}

\title{The End of an Architectural Era\\ for Analytical Databases}

\begin{document}
	
\author{
  Reynold S.~Xin\\
  \affaddr{AMPLab, UC Berkeley}\\
  \affaddr{rxin@cs.berkeley.edu}
}

\maketitle

\abstract{

Traditional enterprise warehouse solutions center around an analytical database system that is monolithic and inflexible: data needs to be extracted, transformed, and loaded into the rigid relational form before analysis. It takes years of sophisticated planning to provision and deploy a warehouse; adding new hardware resources to an existing warehouse is an equally lengthy and daunting task.

Additionally, modern data analysis employs statistical methods that go well beyond the typical roll-up and drill-down capabilities provided by warehouse systems. Although it is possible to implement such methods using a combination of SQL and UDFs \cite{madlib}, query engines in relational databases are ill-suited for these.

The Hadoop ecosystem introduces a suite of tools for data analytics that overcome some of the problems of traditional solutions. These systems, however, forgo years of warehouse research. Memory is significantly underutilized in Hadoop clusters, and execution engine is naive compared with its relational counterparts.

It is time to rethink the design of data warehouse systems and take the best from both worlds. The new generation of warehouse systems should be modular, high performance, fault-tolerant, easy to provision, and designed to support both SQL query processing and machine learning applications.

This paper references the Shark system developed at Berkeley as an initial attempt \cite{shark}.

}

\tinybody{Data warehouse systems should be modular, flexible, easy to provision, and support machine learning. It's time to rethink the system design.}

\bibliographystyle{abbrv}
\bibliography{paper}

\end{document}